\documentclass[sigconf]{acmart}
\AtBeginDocument{%
  }

\copyrightyear{2025}
\acmYear{2025}
\setcopyright{cc}
\setcctype{by-nc-sa}
\acmConference[SC '25]{The International Conference for High Performance Computing, Networking, Storage and Analysis}{November 16--21, 2025}{St Louis, MO, USA}
\acmBooktitle{The International Conference for High Performance Computing, Networking, Storage and Analysis (SC '25), November 16--21, 2025, St Louis, MO, USA}
\acmDOI{10.1145/3712285.3759866}
\acmISBN{979-8-4007-1466-5/2025/11}

\usepackage{bm}
\usepackage{arydshln}
\PassOptionsToPackage{hyphens}{url}

\begin{document}

\title{The First Star-by-star $N$-body/Hydrodynamics Simulation of Our Galaxy Coupling with a Surrogate Model\\
}


\author{Keiya Hirashima}
\email{keiya.hirashima@riken.jp}
\orcid{0000-0002-1972-2674}
\affiliation{
  \institution{\textit{Center for Interdisciplinary Theoretical and Mathematical Sciences (iTHEMS)} \\ RIKEN}
  \city{Wako}
  \country{Japan}
}

\author{Michiko S. Fujii}
\orcid{0000-0002-6465-2978}
\affiliation{
  \institution{\textit{Department of Astronomy} \\
\textit{The University of Tokyo}}
  \city{Tokyo}
  \country{Japan}
}

\author{Takayuki R. Saitoh}
\orcid{0000-0001-8226-4592}
\affiliation{
  \institution{\textit{Department of Planetology and Center for Planetary Science (CPS)} \\
  \textit{Kobe University}}
  \city{Kobe}
  \country{Japan}
}

\author{Naoto Harada}
\orcid{0000-0002-8217-7509}
\affiliation{
  \institution{\textit{Department of Astronomy} \\
  \textit{The University of Tokyo}}
  \city{Tokyo}
  \country{Japan}
}

\author{Kentaro Nomura}
\orcid{0000-0002-2217-2423}
\affiliation{
  \institution{\textit{Preferred Networks, Inc.}}
  \city{Tokyo}
  \country{Japan}
}

\author{Kohji Yoshikawa}
\orcid{0000-0003-0389-5551}
\affiliation{
  \institution{\textit{Center for Computational Sciences} \\
  \textit{University of Tsukuba}}
  \city{Tsukuba}
  \country{Japan}
}

\author{Yutaka Hirai}
\orcid{0000-0002-5661-033X}
\affiliation{
  \institution{\textit{Department of Community Service and Science} \\
  \textit{Tohoku University of Community Service and Science}}
  \city{Sakata}
  \country{Japan}
}

\author{Tetsuro Asano}
\orcid{0000-0002-7523-064X}
\affiliation{
  \institution{\textit{Institut de Ci\`{e}ncies del Cosmos} \\
  \textit{Universitat de Barcelona}}
  \city{Barcelona}
  \country{Spain}
}

\author{Kana Moriwaki}
\orcid{0000-0003-3349-4070}
\affiliation{
  \institution{\textit{Research Center for the Early Universe} \\
  \textit{The University of Tokyo}}
  \city{Tokyo}
  \country{Japan}
}

\author{Masaki Iwasawa}
\orcid{0000-0001-9457-7457}
\affiliation{
  \institution{\textit{Matsue College} \\
  \textit{National Institute of Technology}}
  \city{Matsue}
  \country{Japan}
}

\author{Takashi Okamoto}
\orcid{0000-0003-0137-2490}
\affiliation{
  \institution{\textit{Faculty of Science} \\
  \textit{Hokkaido University}}
  \city{Sapporo}
  \country{Japan}
}

\author{Junichiro Makino}
\orcid{0000-0002-0411-4297}
\affiliation{
  \institution{\textit{Department of Planetology and Center for Planetary Science (CPS)} \\
  \textit{Kobe University}}
  \city{Kobe}
  \country{Japan}
}

\renewcommand{\shortauthors}{Hirashima et al.}

\begin{abstract}
A major goal of computational astrophysics is to simulate the Milky Way Galaxy with sufficient resolution down to individual stars. However, the scaling fails due to some small-scale, short-timescale phenomena, such as supernova explosions. We have developed a novel integration scheme of $N$-body/hydrodynamics simulations working with machine learning. This approach bypasses the short timesteps caused by supernova explosions using a surrogate model, thereby improving scalability. With this method, we reached 300 billion particles using 148,900 nodes, equivalent to 7,147,200 CPU cores, breaking through the billion-particle barrier currently faced by state-of-the-art simulations. This resolution allows us to perform the first star-by-star galaxy simulation, which resolves individual stars in the Milky Way Galaxy. The performance scales over $10^4$ CPU cores, an upper limit in the current state-of-the-art simulations using both A64FX and X86-64 processors and NVIDIA CUDA GPUs. 
\end{abstract}



\maketitle

\section{Overview of the Problem}\label{sec:problem}

\begin{figure}
  \centering \includegraphics[width=9.0cm,clip]{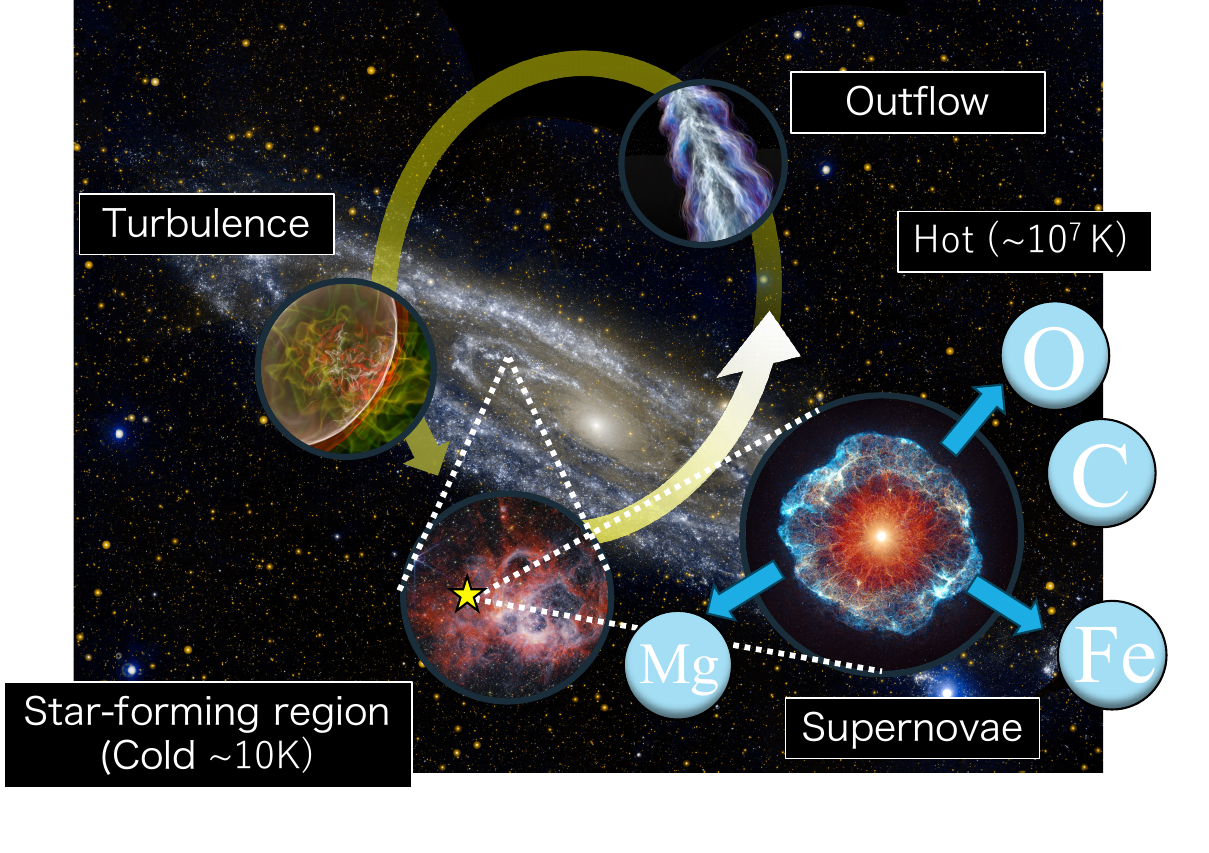}
  \caption{
  Material circulation in a galaxy: Diffuse warm gas loses energy through radiation and conduction and form a disk like structure (galactic disk). Stars form in clouds with low-temperature ($\sim 10$ K) molecular hydrogen in the disk. When massive stars--roughly 10 times the mass of the Sun--reach the end of their lifetimes, they explode as supernovae, generating extremely hot gas ($\sim10^7$ K). These explosions inject both energy and heavy elements, such as carbon (C), oxygen (O), magnesium (Mg), and iron (Fe) into the surrounding interstellar gas and induce turbulence. A part of these materials is ejected as outflow and eventually fall back to the galactic disk, where forms the next generation stars. These enriched materials finally forms planets like the Earth and lives like us.
  (credit: NASA/JPL-Caltech, ESA, CSA, STScI). 
}
\label{fig:gas_dynamics}
\end{figure}

Chemical elements of the universe are synthesized mostly in stars, except for hydrogen and helium, which were formed just after the Big Bang. Elements synthesized inside stars spread via supernova explosions, which typically release the energy of $10^{51}$ erg. These elements mix with the surrounding interstellar matter, mostly hydrogen, and form new generations of stars. This cycle continues for 10 Gyr ($=10^{10}$ yr) inside galaxies as illustrated in Figure ~\ref{fig:gas_dynamics} and finally results in the formation of the Earth and lives on it. Such a long time evolution of the universe can be studied using numerical simulations. 

Galaxies are stellar systems composed of a few hundred billion of stars and interstellar gas (baryon) embedded in a dark matter (DM) halo with a mass of 20--100 times more than the baryon. The Sun is one of $>10^{11}$ stars of the Milky Way (MW) Galaxy. The dynamics of galaxies is governed by gravity. Gravity gathers DM to be bound. In such bound DM halos, the gas component sinks into the center of DM halos and forms stars. If the gas has angular momentum, the gas and stars form a rotationally supported galactic disk. The MW Galaxy is one of these disk galaxies. Stars are known to follow a mass spectrum. Massive stars more than about 10 times solar masses ($M_{\odot}$) are only a few percent of all stellar populations but play important roles by their radiative heating to interstellar gas and supernova explosions at the end of their lifetimes. Supernovae (SNe) inject energy and materials created inside stars into their surrounding gas and create turbulence and outflow. These complicated, nonlinear phenomena must be solved with numerical simulations. 

$N$-body/smoothed-particle hydrodynamics (SPH) simulations are widely used for galaxy simulations. Stars and DM are modeled as $N$-body particles contributing as gravitational sources. In contrast, interstellar gas is modeled with SPH particles, and the gas distribution is realized with the distributions smoothed by the kernel radius, which is typically the size of 100 gas SPH particles.

The DM halo of the MW Galaxy extends to 200,000 pc (1\,pc $= 3\times 10^{16}$\,m), while SN shell scale is a few pc. The highest temperature of the gas reaches $10^7$\,K, but the star-forming molecular gas is $\sim 10$\,K. The timescale of expanding SN shell is years, but the timescale of the galactic disk rotation is $10^8$ years. Thus, the physical scales of galaxies spread over a range of 5--6 orders, and therefore, performing high-resolution galaxy simulations is technically challenging.
So far, the maximum number of particles used in state-of-the-art simulations is limited to less than one billion (see Table~\ref{tab:pastsims_iso}). Because the total mass of the MW Galaxy is an order of $10^{12}$ $M_{\odot}$ \cite{McMillan2017}, the highest mass resolution was $400 M_{\odot}$ for star and gas and $\sim 10^4 M_{\odot}$ for DM \cite{Richingsetal2022}.  
For small galaxies with 1/100 mass of the MW Galaxy, the resolution reached $1 M_{\odot}$ \cite{Steinwandel+24a}. The total number of particles is also less than one billion. Thus, one billion particles is a barrier we have to break through.

The bottleneck in galaxy simulations arises from the need for small timesteps in localized regions with increased resolution.
The most severe timestep condition is the Courant-Friedrichs-Lewy (CFL) condition, which limits the timestep of hydro components (e.g., gas). In this condition, the required timestep is expressed as the scale of a fluid element over the sound speed, and particularly, it becomes extremely small in the dense hot gas around SNe. 
The timestep is expected to be nearly proportional to the mass of the particle, $m$, ($dt_{\rm CFL} \propto \rho/m^{1/3} \propto m^{5/6}$, where $\rho$ is the gas density). Adopting the typical sound speed of an SN region ($1000~{\rm km~s^{-1}}$), the required timestep becomes an order of 100 yr for $1M_{\odot}$ resolution, while the simulation time we want to integrate is $10^9$ years.

Strong scaling gets worse for more than a few thousand CPU cores \cite{Springel+2020,Hopkins+2018}. 
In such recent galaxy simulations, individual or hierarchical timestep methods are often adopted \cite{McMillan1986,HernquistKatz1989}. In this method, each particle has its own timestep and is updated only when an integration is required. 
The computational efficiency tends to decrease when the fraction of particles to be updated is small because inter-process communications must be done at each timestep. 
For example, we need to predict the positions and other physical quantities of all particles and construct a Barnes-Hut octree\cite{BarnesHut1986} structure for the force calculation. These processes consume time for communication that is comparable to that required for updating all particles. As a result, smaller timesteps worsen efficiency in high-resolution simulations, even when individual or hierarchical timestep methods are employed. These small timesteps worsen the parallelization efficiency because a small number of particles can be integrated in one step. The use of GPUs also faces the same problem. 
Thus, we need to avoid small hierarchical timesteps to improve the time-to-solution and scalability. 
In this paper, we break the billion-particle barrier using our new integration scheme coupled with a surrogate model.

\section{Current State of the Art}

\begin{table*}[htb]
	\centering
	\caption{List of state-of-the-art hydrodynamics simulations of isolated disk galaxies. From left to right, columns show the authors of the simulation papers, number of gas particles ($N_{\rm{gas}}$), gas particle mass ($m_{\rm{gas}}$), number of star particles ($N_{\rm{star}}$), star particle mass ($m_{\rm{star}}$), number of DM particles ($N_{\rm{DM}}$), total mass ($M_{\rm tot}$), total number of particles ($N_{\rm tot}$), used code, and references.}
	\label{tab:pastsims_iso}
	\begin{tabular}{lccccccccc} 
		\hline
		Paper & $N_{\rm{gas}}$ &$m_{\rm{gas}}$ [$M_{\odot}$]& $N_{\rm star}$& $m_{\rm{star}}$ [$M_{\odot}$] & $N_{\rm DM}$ & $M_{\rm tot}$ [$M_{\odot}$] & $N_{\rm tot}$& Code & Ref.\\
		\hline \hline
        Hu et al. (2017) & 10$^{7}$ & 4 & $10^{7}$ & 4 & $4 \times 10^{6}$ &  $2 \times 10^{10}$ &$2.4\times 10^{7}$& GADGET-3 &\cite{Hu+2017}\\
        
        Smith et al. (2018) & $1.9 \times 10^{7}$ & 20 & $10^{5}$ & 20 & $10^{5}$  & $10^{10}$  & $2.0\times 10^{7}$ &AREPO&\cite{Smith+18}\\
        Smith et al. (2018) Large & $1.9 \times 10^{7}$  & 200 & $10^{5}$ & 200 & $10^{5}$ &  $10^{11}$  &$2.0\times 10^{7}$& AREPO &\cite{Smith+18}\\
        
		Smith et al. (2021) & $3.4 \times 10^{6}$  & 20 & $4.9 \times 10^6$& 20 & $6.2 \times 10^{6}$ &  $10^{10}$  &$2.0\times 10^{7}$& AREPO &\cite{Smith+21}\\
        
        Richings et al. (2022) &  $10^{7}$  
        & 400 & $3\times10^7$ & 400 & $1.6\times10^8$ & $10^{12}$  &$2.0\times 10^{8}$ & GIZMO & \cite{Richingsetal2022}\\
        
        Hu et al. (2023) & $7 \times 10^{7}$  & 1 & $10^{7}$ & 1 & $10^{7}$ &  $10^{10}$  &$2.4\times 10^{7}$ & GIZMO & \cite{Hu+23b}\\
        
        Steinwandel et al. (2024) & $10^8$ & 4 & $5 \times 10^8$ & 4 & $4 \times 10^{7}$ & $2\times10^{11}$&$6.4\times 10^{8}$& GADGET-3 & \cite{Steinwandel+24a} \\ 
    
        \hline
        
        This work & $4.9\times10^{10}$ & 0.75 & $7.2\times10^{10}$ & 0.75 & $1.8\times10^{11}$ & $1.2\times10^{12}$ & $3.0\times 10^{11}$ & ASURA & -\\
		\hline
	\end{tabular}

\end{table*} 

\begin{figure} 
  \centering 
  \includegraphics[width=9.0cm,clip]{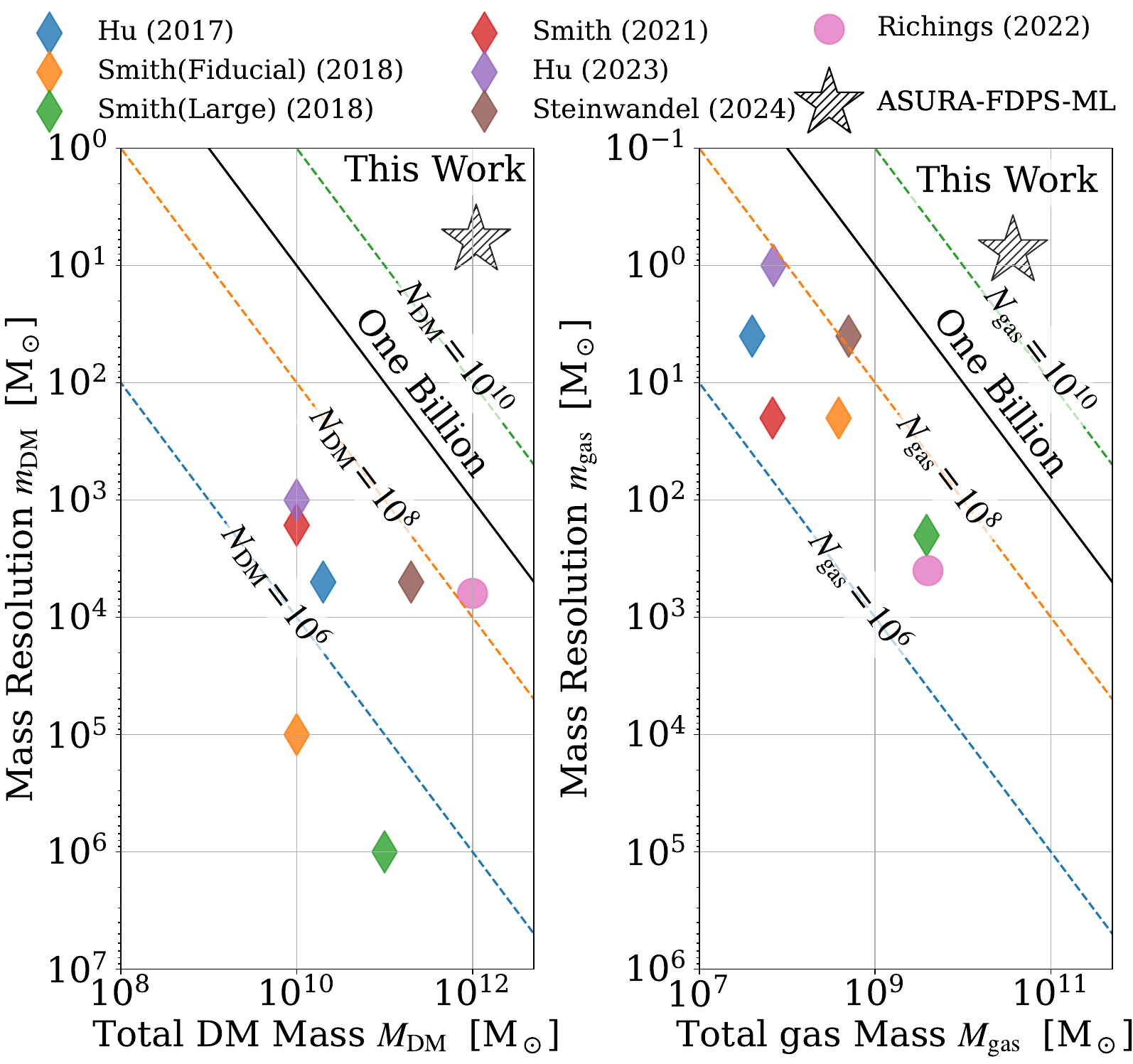}

  \caption{The total mass of the system and the resolution of the DM (left) and gas (right) particles of the current state-of-the-art simulations listed in Table \ref{tab:pastsims_iso}. 
  Diagonal dotted lines represent the constant number cases of $N_{\rm DM} (N_{\rm gas})$ = $10^6, 10^8$ and $10^{10}$ for a system. The black-solid line indicates the billion-particle barrier.} 
  \label{fig:Res} 
\end{figure}

Even in the current state-of-the-art galaxy simulations, the number of particles is limited to $<10^9$ as mentioned in Section \ref{sec:problem}. Therefore, the current state-of-the-art simulations are categorized as either MW-size galaxies with low mass resolution ($>100 M_{\odot}$) or smaller galaxies with star-by-star resolution as summarized in Table~\ref{tab:pastsims_iso}.
Figure~\ref{fig:Res} shows these simulations with respect to mass resolution.
The highest resolution of a MW-size galaxy simulation was performed in Richings et al. (2022) \cite{Richingsetal2022} using $\sim10^7$ particles for gas and stars and $10^8$ particles for DM. This setup results in a mass resolution of $400 M_{\odot}$ for star and gas particles, which is two orders of magnitude lower than a realistic stellar mass ($1M_{\odot}$).
The other simulations with a higher resolution modeled $1/10$ or $1/100$ smaller galaxies that are similar to dwarf galaxies orbiting around the MW Galaxy. For such smaller galaxies, Hu et al. (2023)\cite{Hu+23} resolved down to $1M_{\odot}$ using $\sim 10^8$ particles for the gas and stars. 
Steinwandel et al. (2024)\cite{Steinwandel+24a} simulated a galaxy with a $1/10$ size of the MW Galaxy. The gas and stellar mass resolution was $4 M_{\odot}$, which is nearly resolving individual stars.

As shown in Table~\ref{tab:pastsims_iso}, these state-of-the-art simulations have been done by three simulation codes: GIZMO, AREPO, and GADGET.
The GADGET series\footnote{https://wwwmpa.mpa-garching.mpg.de/gadget4/} \cite{Springel2005, Springel+2020} comprises tree-based force evaluation methods (the tree code and fast multipole method) and SPH for compressive fluid. GIZMO\footnote{http://www.tapir.caltech.edu/~phopkins/Site/GIZMO.html}\cite{Hopkins+2018}, derived from GADGET, implements a recently developed mesh-free method for hydrodynamics that offers greater accuracy than SPH. AREPO\footnote{https://arepo-code.org/}\cite{Arepo2020} represents a new class of astrophysical simulation codes, using the finite-volume method for fluid dynamics and Voronoi tessellation to define dynamically evolving astrophysical structures. Its force-evaluation approach remains similar to that of GADGET.
With any of these codes, the highest resolution is similar, i.e., star-by-star for $<1/10$ MW-sized galaxies and $>100 M_{\odot}$ for MW-like galaxies.

The billion-particle barrier is not only for isolated galaxy simulations; galaxy formation simulations in a cosmological context also have the same barrier. 
The largest number of gas particles in a larger scale simulation is $10^{8}$ \cite{Applebaum+2021}, and the highest mass resolution is $5\times 10^3~M_{\odot}$ for DM and $8\times 10^2~M_{\odot}$ for baryon (gas and stars) \cite{Auriga2}. 

Without gas, the limit of the maximum number of particles is relaxed. 
B\'{e}dorf et al. (2014) \cite{Bedorfetal2014}, one of finalists for the 2014 Gordon-Bell Prize, performed the largest simulation of a disk galaxy ever achieved (the number of particles was $\sim10^{11}$), in which a MW-sized galaxy that consists of DM halo and stellar disk was modeled with particles. Practically, several billion particles are used for scientific papers \cite{Fujii+2019}. 
In the past, Gordon-Bell winners with $N$-body simulations were all without gas, such as Ishiyama et al. (2012 Gordon-Bell Prize)\cite{Ishiyama+2012}. 
These gravity-only simulations have no constraint from the CFL condition, allowing them to have longer timesteps than those in hydrodynamics simulations. 
Thus, performing high-resolution $N$-body/SPH simulations of galaxies using the recent world's largest supercomputers is a big challenge.

\section{Innovations Realized: Deep Learning Working with Simulations}

\subsection{Overview}

\begin{figure*}
  \centering \includegraphics[width=14.0cm,clip]{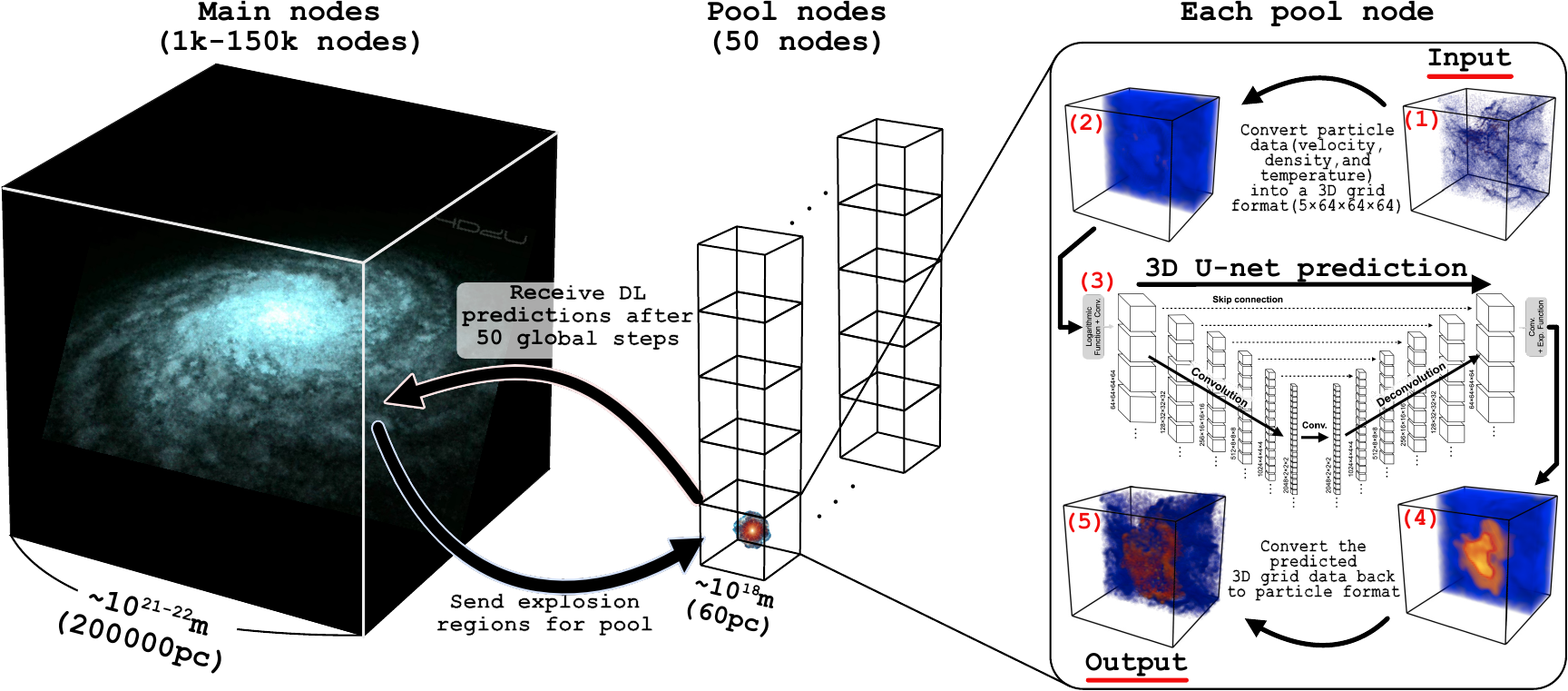}
  \caption{
                  Schematic illustration of our simulation method. The main nodes integrate the entire region of a galaxy using a shared timestep ($\Delta t_{\rm global}$) with a large number of computational nodes (i.e., $1~{\rm k} \sim 150~{\rm k}$ nodes). Upon detecting SN events, it sends the affected regions to an available pool node. This pool node then uses a pre-trained neural network to predict the 3D evolution of these SN regions. The prediction process is carried out independently from the simulation performed by the main nodes. Every 50 global timesteps, the predicted particle data is sent back to the main nodes. To handle the continuous processing of SN events, the system maintains a set of 50 pool nodes, corresponding to the 50-step interval between updates. \textcopyright 2010 Takaaki Takeda, Junichi Baba, Takayuki Saitoh, 4D2U Project, NAOJ.}
  \label{fig:overview}
\end{figure*}

The bottleneck of state-of-the-art galaxy simulations is caused by small timesteps required for small-scale phenomena such as supernova explosions. We therefore developed a scheme to bypass the time evolution of supernova shells using a surrogate model instead of integrating them. 
Here, we briefly describe an overview of our scheme. The details of the scheme and validation are summarized in \cite{Hirashima2025ApJ}.
Figure~\ref{fig:overview} shows a schematic picture of our scheme. 
We split the MPI communicator into two: one is for normal $N$-body/SPH integration, and the other is for predicting the particle distribution using deep learning (DL). We call the former `main nodes' and the latter `pool nodes.' The number of pool nodes is small ($<50$) compared to the main nodes. 

Once an SN is detected from the stellar evolution model we adopt, the SPH particles in a cube with a side length of 60 pc around the SN are sent to a pool node. The DL predicts the distribution of gas after 100,000 years in a pool node and sends the SPH particle data back to the main node(s). During this process, the main nodes continue integration without knowing the SN results. If new SNe occur at the next step, the particles around it are sent to another pool node.
Thus, the integration of the galaxy using the main nodes and the prediction of the SN region with DL using the pool nodes fully overlap. 
Hereafter, we describe the details of our method.

\subsection{Integration of the entire galaxy with deep learning}

We integrate the entire galaxy with the second-order leapfrog scheme. 
The integration of one step using a leapfrog scheme with a shared timestep generally proceeds as follows: (1) Initial velocity change for $1/2 \Delta t$, (2) drift all particles, (3) evaluate force, (4) velocity change for $1/2 \Delta t$, (5) star formation and feedback etc., (6) recalculate hydro force and kernel size, and (7) determine the next timestep.

In this general implementation, when an SN explosion occurs, the timestep for the next step is shortened. 
In our new scheme, we identify SNe exploding in the next step, send the SPH particles around them to one of the pool nodes, and predict the shell expansion using DL in the pool node (see Figure~\ref{fig:overview}). 
The entire procedure is:
\begin{enumerate}
    \item Identify stars exploding between the current time $t$ and $t+\Delta t _{\rm global}$.
    \item Pickup particles in the (60\,pc)${^{3}}$ box around the exploding star and send them to a pool node, which performs DL prediction of SNe that occur in this step.
    \item Calculate the first velocity change, drift, force evaluation, and the second velocity change in the main nodes without adding any feedback energy.
    \item Receive particles from the pool node and replace the particles with them in the main nodes referring to the particle IDs. 
    \item Decompose the domain and exchange particles.
    \item Create new stars and calculate cooling and heating. 
    \item Recalculate hydro force, etc., after changing the internal energy.
    \item Go back to step 1.
\end{enumerate}
In this method, we can adopt a fixed global timestep $\Delta t _{\rm global}$.

The pool node gives the particle distribution 0.1\,Myr ($=10^5$\,yr) after the explosion using DL prediction.
As we have multiple pool nodes, we can set a global timestep smaller than the timestep for the DL prediction. 
If $\Delta t_{\rm global}=$2,000\,yr, for example, we adopt 50 pool nodes. The pool nodes predict the particle distribution after $50\Delta t_{\rm global}$ and send the distribution back to the main nodes after $50\Delta t_{\rm global}$.

\subsection{Deep-learning surrogate model}

We developed a DL model to predict the expansion of the SN shell. 
Specifically, our model predicts the distributions of five physical quantities of gas: density, temperature, and velocity in three directions.
To prepare training data, we conduct SN explosion simulations with a gas particle resolution of 1 $\rm M_\odot$, and obtain the gas distributions just before the explosion and after 0.1 Myr. 
As initial conditions, we use density fields disturbed by turbulent velocity fields that follow $\propto v^{-4}$, which imitate environments of star-forming regions in MW-like galaxies.

We employ a U-Net architecture \cite{Ronneberger+15} for our DL model. Our model consists of a series of three-dimensional convolutional layers (Figure \ref{fig:overview}). Before applying convolutions, the particle data should be pre-processed into structured grid data. We do this by mapping gas particles into voxels using the SPH kernel convolution and the Shepard algorithm \cite{Shepard+68}.
Similar mapping schemes have been used in several machine learning applications for particle simulations \cite{Jamieson+23,Hirashima+23a,Hirashima+23b,Chan+24}.
The data cube is cut out so that the location of the SN explosion is at its center.
The obtained data cube has a side length of 60 pc and is composed of $64^3$ voxels.
When we obtain an output of structured grid data from the machine, we convert it back to particle data 
using Gibbs sampling, which is one of the Markov chain Monte Carlo methods.
Mass conservation is ensured by making the number of created particles the same as the number of particles in the input data.

A general and crucial problem when applying a DL model to compressible hydrodynamics data is the dynamical range of physical quantities, which spans several orders of magnitude. 
For instance, the temperature changes by as much as six orders of magnitude in a SN explosion.
This makes it difficult for a machine to handle the SN simulation data.
To avoid such a problem, we take the logarithm of the physical quantities before inputting the U-Net. For the three velocity fields, we divided each of them into two data cubes, one for pixels with positive velocities and another for those with negative velocities, and take the logarithm of their absolute values. 
We thus input a total of eight data cubes into the machine.

Our model is implemented using Keras and TensorFlow \cite{tensorflow}
and trained using a single NVIDIA A100 Tensor Core GPU.
We perform training with a batch size of 1 with the mean squared error between the true (simulated) and predicted physical quantities.
We used the model trained for 100 epochs hereafter because the validation error converged and stabilized around 100 epochs.
ADAM optimizer \cite{Kingma+2014} is adopted with a learning rate of $10^{-6}$.
While DL models are generally trained and used on GPUs with Python libraries, if we incorporate a model optimized for GPUs with a numerical simulation that runs on CPUs, the data transfer between GPUs and CPUs could be a new bottleneck. 
To avoid this, we abandon using GPUs for inference; we implement the code for DL inference with C++ and optimize it for CPUs by exploiting Open Neural Network Exchange (ONNX) \cite{onnxruntime} for the x64 architecture and SoftNeuro \cite{Hilaga+2021} for the Arm architecture.

\begin{figure}

\includegraphics[width=8.5cm]{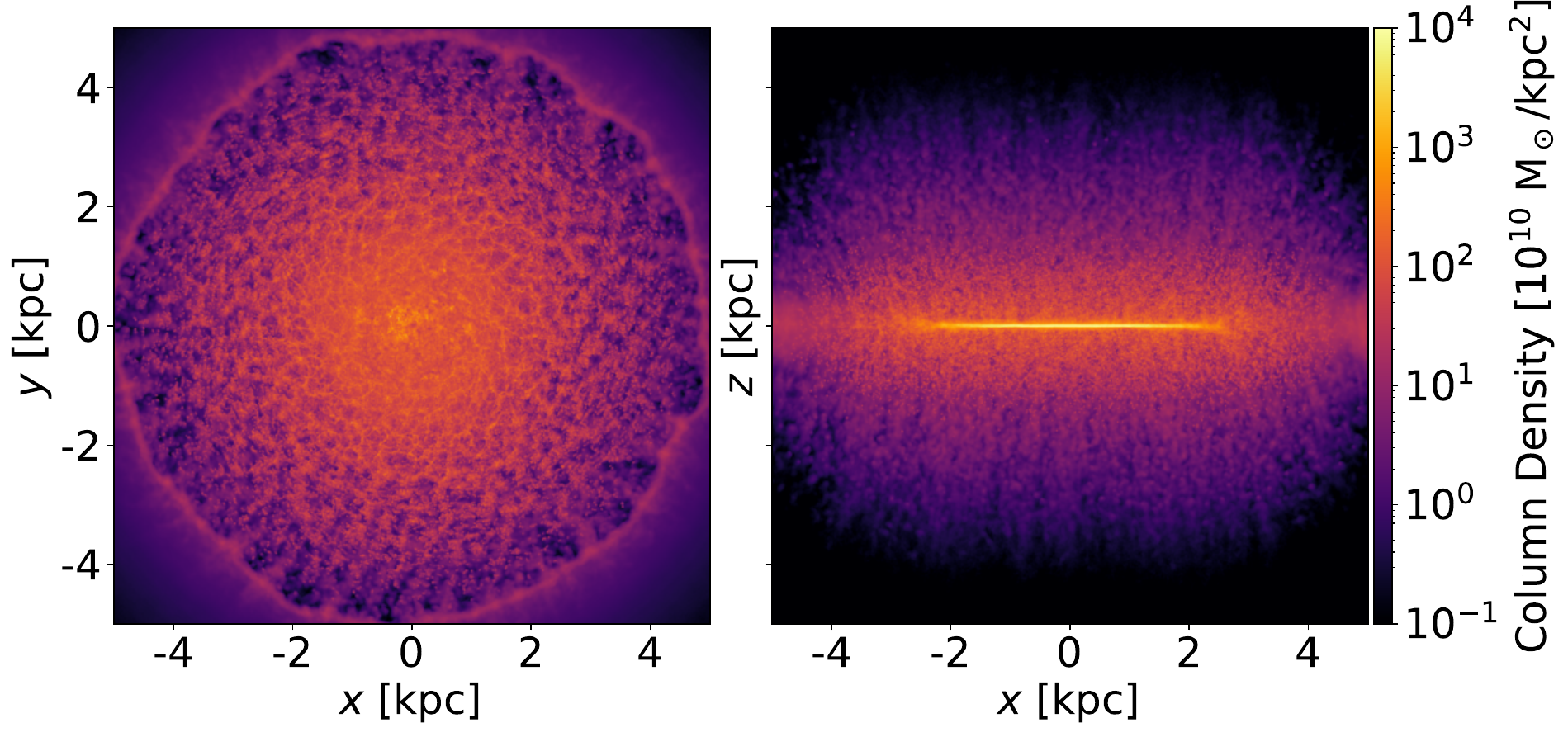}
\caption{Snapshots of gas distribution of the galactic disks integrated with our new scheme with DL surrogate model. The right and left panels show surface density for the face-on ($x-y$ plane) and edge-on ($x-z$ plane), respectively.}
\label{fig:snapshot}
\end{figure}

In Figure~\ref{fig:overview}, we present an example of machine learning prediction. We confirmed that the prediction is better than low-resolution simulations by comparing the total energy and momentum \cite{Hirashima2025ApJ}. We also confirmed the accuracy of our new scheme using some indicators obtained from the global structures of galaxies, such as star formation rates and mass loading factors \cite{Hirashima2025ApJ}. As shown in Figure~\ref{fig:snapshot}, the new scheme with the surrogate model cannot be distinguished from conventional simulations, which integrate all particles. 
This scheme has also been validated through direct comparison with results from conventional numerical simulations\cite{Hirashima2025ApJ}.
We also confirmed that the probability distribution functions of gas density and temperature are reproduced with the surrogate model for SNe \cite{Hirashima2025ApJ}. 
We emphasize that such a complex morphology cannot be reproduced with any other analytical (sub-grid) method.

\subsection{Framework for Developing Particle Simulators}\label{sec:FDPS}

Framework for Developing Particle Simulators (FDPS)\footnote{https://jmlab.jp/fdps/} is a general-purpose, high-performance library for particle simulations. We used this library, adding some modifications for massive parallel computing with $>10,000$ MPI processes. 

FDPS has functions necessary for particle-particle interaction calculations using a treecode\cite{BarnesHut1986}, in which particles are assigned to a tree structure and the calculation cost becomes $O(N\log N)$ instead of $O(N^2)$. FDPS provides functions for domain decomposition, particle exchange, tree construction, local essential tree (LET) exchange, and user-defined interaction calculation using the tree. 

The bottleneck is the all-to-all communication. 
In galaxy simulations, domain decomposition and the following particle and local tree (LET) exchanges require communication among entire MPI processes. We implemented the algorithm whose time complexity is $O(p^{1/3})$, where $p$ is the number of MPI processes \cite{Iwasawaetal2019}.
We used the 3D {\tt MPI\_Alltoallv} algorithm, in which three MPI communicators are defined and they match the 3D torus node configuration and domain decomposition.
When {\tt MPI\_Alltoallv} is called, the 3D {\tt MPI\_Alltoallv} algorithm calls {\tt MPI\_Alltoallv} three times for each MPI communicator.
This algorithm reduces the number of nodes joining one {\tt MPI\_Alltoallv} operation, and avoids the global communication of all the main nodes.
Such MPI parallelization is realized inside the FDPS library.
FDPS is also designed for multiple platforms and is GPU compatible. 

\subsection{Tuning of particle-particle interaction kernels: PIKG}
Besides the timestep problem, particle-particle interaction calculations are the heaviest and generally become bottlenecks in galaxy simulations. For example, at every timestep, a particle needs gravitational force from all the other particles. 
Equation~\ref{eq:interation} gives the definition of the particle-particle interaction for gravity:
\begin{equation}
  \bm F_{{\rm grav},ij} = - G \dfrac{m_i m_j}
      {(r_{ij}^2+\epsilon_i^2 +\epsilon_j^2)^{3/2} } \bm r_{ij},
  \label{eq:interation} 
\end{equation}
where $\bm r_i$, $m_i$, and $\epsilon_i$ are the position,
mass and the softening parameter of particle $i$, and $G$ is the gravitational
constant, respectively, and $\bm r_{ij} = \bm r_i - \bm r_j$ and $r_{ij} = \| \bm r_{ij} \|$.
The value of the softening parameter depends on both the resolution (particle mass) and the types
of particles (DM/gas/stars).  
Tuning the particle-particle interaction kernels is the key to the optimization of galaxy formation simulations.

To solve this problem, we have developed an automatic Particle-particle Interaction Kernel Generator (PIKG{\footnote {\url{https://github.com/FDPS/PIKG}}}), which takes the high-level description of interaction kernels written in a simple DSL and generates code in many different forms, including intrinsics for the ARM SVE architecture. 
The generated code for A64FX using ARM SVE intrinsics is about 500
lines. 
In this code, (1) automatic conversion between the structure of
arrays and arrays of structure, (2) loop unrolling, and (3) loop fission (necessary for Fujitsu A64FX) are applied.

For efficient computation, we employed the piecewise polynomial approximation (PPA) for the computation of the kernel function in SPH kernels. In PPA, the domain of the target function is divided into $m$ subdomains.
The function in each subdomain is approximated by the $n$th-order polynomials. Thus, $m(n+1)$ coefficients of the polynomials are needed.
We used Sollya \cite{Chevillard2010Sollya} for computing the minimax polynomials to approximate the target function in each subdomain.
The approximated function of section $k$ is
\begin{eqnarray}
 f_{\mathrm{PPA}}^{\mathrm{app}}(x;k) = \sum_{l=0}^n a_{k,l}(x - k d)^l
\end{eqnarray}
where $a_{k,l}$ is the coefficient of the $l$th term in the polynomial of section $k$, and $d$ is the length of each subdomain.
In modern SIMD CPU environments such as ARM SVE and AVX-512, PIKG utilizes a table lookup function, which enables SIMD registers to accommodate table coefficients that bring fast calculation of the polynomials. 

\section{How Performance Was Measured}

\subsection{System and Environment}
We have performed our numerical simulations on three supercomputers with different architectures. 
\subsubsection{Fugaku}
Fugaku supercomputer consists of 158,976 computational nodes, each of which has a Fujitsu A64FX processor. 
The A64FX processor has 48 compute cores,
and the total memory per node is 32\,GB.
The theoretical peak performance for a single processor running at 2.0 GHz is 6.144\,TF for single precision and 3.072\,TF for double precision. 
TofuD, a six-dimensional mesh/torus network, is used to connect the nodes.
We measured the performance with up to 152,064 nodes, 95\% of the entire system. We run one MPI process per node and 48 OpenMP threads per MPI process to relax the memory limitations.

\subsubsection{Flatiron, Rusty cluster, genoa node}
The genoa node of the Rusty cluster at Flatiron Institute consists of 432 nodes, each of which has two genoa (AMD EPYC™ 9474F) processors. Each processor has 48 compute cores and 48 threads.
The total memory per node is 1.5 TB.
The theoretical peak performance for a single processor running at 4.1 GHz is 6.298\,TF for single precision and 3.149\,TF for double precision.
The calculation nodes are connected with InfiniBand.
We measured the performance with up to 193 nodes, 45\% of the entire system. We run 48 MPI processes per node and 2 OpenMP threads per MPI process.

\subsubsection{Miyabi}
Miyabi (Miyabi-G) consists of 1,120 nodes, each of which has one NVIDIA Grace CPU (72 cores 3.0GHz) and NVIDIA Hopper H100 GPU (66.9TF).
The CPU and GPU are connected NVLink-C2C with NVIDIA GH200 Grace-Hopper Superchip.
The memories of CPU and GPU are 120 GB and 96 GB, respectively.
The theoretical peak performance of the entire system is 78.8 PF for double precision.

\subsection{Model}
\label{sec:model}

\begin{figure}
\centering
\includegraphics[width=6.5cm]{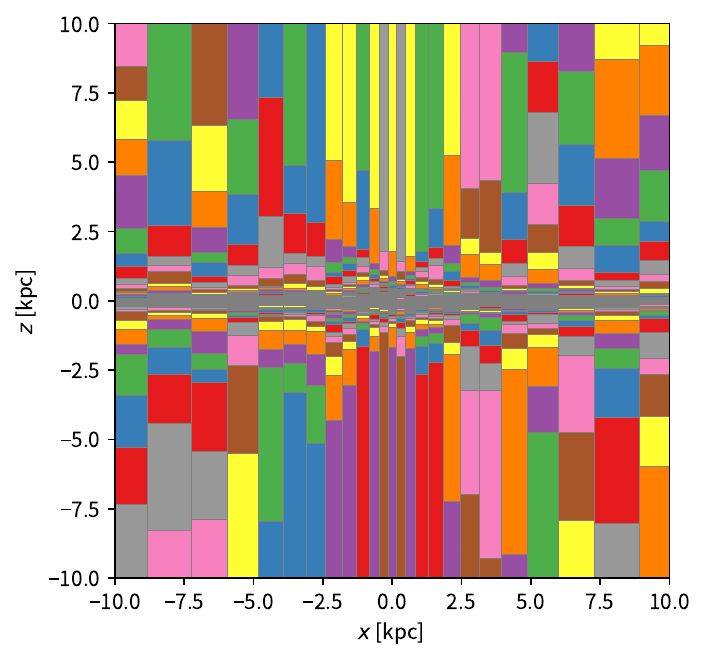}
\caption{An example of the domain decomposition sliced at $y=0$.}
\label{fig:DomainCoord}
\end{figure}

We generated initial conditions using Action-based Galaxy Modelling Architecture (AGAMA) \cite{Vasiliev2019}\footnote{https://github.com/GalacticDynamics-Oxford/Agama} modified for parallel generation for each domain\footnote{https://github.com/tetsuroasano/Agama}.
The parameters are adjusted to reproduce the MW Galaxy\cite{McMillan2017}. 
The model is composed of three components: DM, stars, and gas. The DM distributes in a broken power-law. Inside this DM halo, stars and gas distribute a rotating disk. The halo is mainly composed of DM, but some stars and gas are also distributed. 
The total mass of each component is $1.1\times 10^{12} M_{\odot}$ for DM, $5.4\times10^{10} M_{\odot}$ for stars, and $1.2\times 10^{10} M_{\odot}$ for gas. We refer to this model as Model MW. 
We generated the initial particle distribution for each domain at the beginning of the simulation. DM and stellar particles are sampled from distribution functions. The equilibrium gas disk is generated with the potential method\cite{Wang2010}. 
The mass resolution is summarized in Table.~\ref{tab:runs}.

We note that the distribution of particles is highly concentrated in the center. The halo radial density follows a broken power-law function, and in the central region, the density increases with $\propto r^{-1}$, where $r$ is the distance from the galactic center. The disk surface density exponentially increases toward the galactic center. The scale height of the disk is only $\sim 10$\,\% of the scale length. Therefore, the distribution of particles is highly concentrated in the center and disk plane. 
Figure \ref{fig:DomainCoord} shows an example of the domains assigned to each node. As shown in this plot, the domains are highly concentrated in the center and the mid-plane, and the shapes of the domains are often very narrow. 
We also utilized a disk galaxy but $1/10$ mass (Model MW-small) and $1/100$ mass (Model MW-mini).

\subsection{Measurement Methodology}
We inserted {\tt MPI\_Barrier} and {\tt MPI\_Wtime} before and after critical routines in the main nodes to measure timing results.
For flop measurements, we used {\tt fapp} for Fugaku. For the other systems, we counted the number of interactions that evaluate gravity and hydro force, multiplied the number of operations of those interactions, and finally divided them by the measured timings. 
The numbers of operations are summarized in Table \ref{tab:PIKGgravity}.

Positions and velocities of particles are stored in double-precision variables to handle a wide range of orders of more than five magnitudes. However, the relative accuracy necessary for the interaction calculation is single precision. Therefore, we implemented a mixed-precision scheme. When we calculate force from a group of particles (particles in the interaction list) to another group of particles, the positions and velocities of the particles are first converted to the values relative to the representative value of the particles that receive the force and then converted to single precision. In this method, we can maintain sufficient accuracy and double-precision resolution while using single-precision calculations for the most time-consuming interaction calculation. 


\begin{table*}
\centering
\caption{List of runs. From left to right, columns show run name, the number of the main nodes ($N_{\rm{node}}$), the mass of one DM particle ($m_{\rm{DM}}$), the number of DM particles ($N_{\rm{DM}}$), the mass of one star particle ($m_{\rm star}$), the number of star particles ($N_{\rm star}$), the mass of one gas particle ($m_{\rm{gas}}$), the number of gas particles ($N_{\rm{gas}}$), and the total number of particles ($N_{\rm tot}$) per node.} 
\label{tab:runs}
\begin{tabular}{lrrrrrrrrr} 
\hline
Run &  $N_{\rm{node}}$ & $m_{\rm{DM}}$ & $N_{\rm{DM}}$ & $m_{\rm star}$ & $N_{\rm star}$ &  $m_{\rm{gas}}$  & $N_{\rm{gas}}$& $M_{\rm tot}$ & $N_{\rm tot}$/$N_{\rm node}$ \\
 &  &  [$M_{\odot}$] & &  [$M_{\odot}$] &  &  [$M_{\odot}$] &  & [$M_{\odot}$] &\\
\hline
weakMW2M & 148896--128 & 6.0 & $1.8\times 10^{11}$ & 0.75 & $7.2\times10^{10}$ & 0.75 & $4.9\times10^{10}$ & $1.2 \times 10^{12}$& $2\times 10^{6}$ \\

\hdashline

weakMW\_rusty & 193--11 & 7.7 & $1.4\times 10^{11}$ & 0.96 & $5.5\times10^{10}$ & 0.96 & $3.8\times10^{10}$ & $1.2 \times 10^{12}$& $1.2\times 10^{9}$ \\

\hline

strongMW & 148896--67680 & 11.7 & $9.3\times 10^{10}$& 1.4 & $3.7\times10^{10}$ & 1.4 & $2.6\times10^{10}$ & $1.2\times10^{12}$& $1.0$--$2.3\times 10^{6}$ \\

strongMWs & 40608--4096 & 4.0 & $2.8\times 10^{10}$ & 0.5 & $1.2\times10^{10}$ & 0.5 & $7.5\times10^{9}$ & $1.2\times10^{11}$& $1.2$--$12.0\times 10^{6}$ \\

strongMWm & 1024--128 & 12.0 & $1.4\times 10^{9}$ & 1.5 & $3.7\times10^{8}$ & 1.5 & $3.4\times10^{9}$ & $1.8\times10^{10}$& $2.1$--$16.0\times 10^{6}$ \\

\hdashline

strongMW\_rusty & 193--43 & 36.0 & $3.0\times 10^{10}$ & 4.5 & $1.2\times10^{10}$ & 4.5 & $8.4\times10^{9}$ & $1.2\times10^{12}$& $2.6$--$11.9\times 10^{8}$ \\

strongMWs\_rusty & 43--11 & 166 & $6.5\times 10^{9}$ & 21 & $2.6\times10^{9}$ & 21 & $1.8\times10^{9}$ & $1.2\times10^{12}$& $2.5$--$99.4\times 10^{8}$ \\

\hline
MW\_miyabi & 1024 & 87.9 & $1.2\times 10^{10}$ & 11 & $5.0\times10^{9}$ & 11 & $3.4\times10^{9}$ & $1.2\times10^{12}$& $2.0\times 10^{7}$ \\

\hline
\end{tabular}
\end{table*} 

\section{Performance Results}

\subsection{Scalability}
\begin{figure*}
\centering
\includegraphics[width=7.5cm]{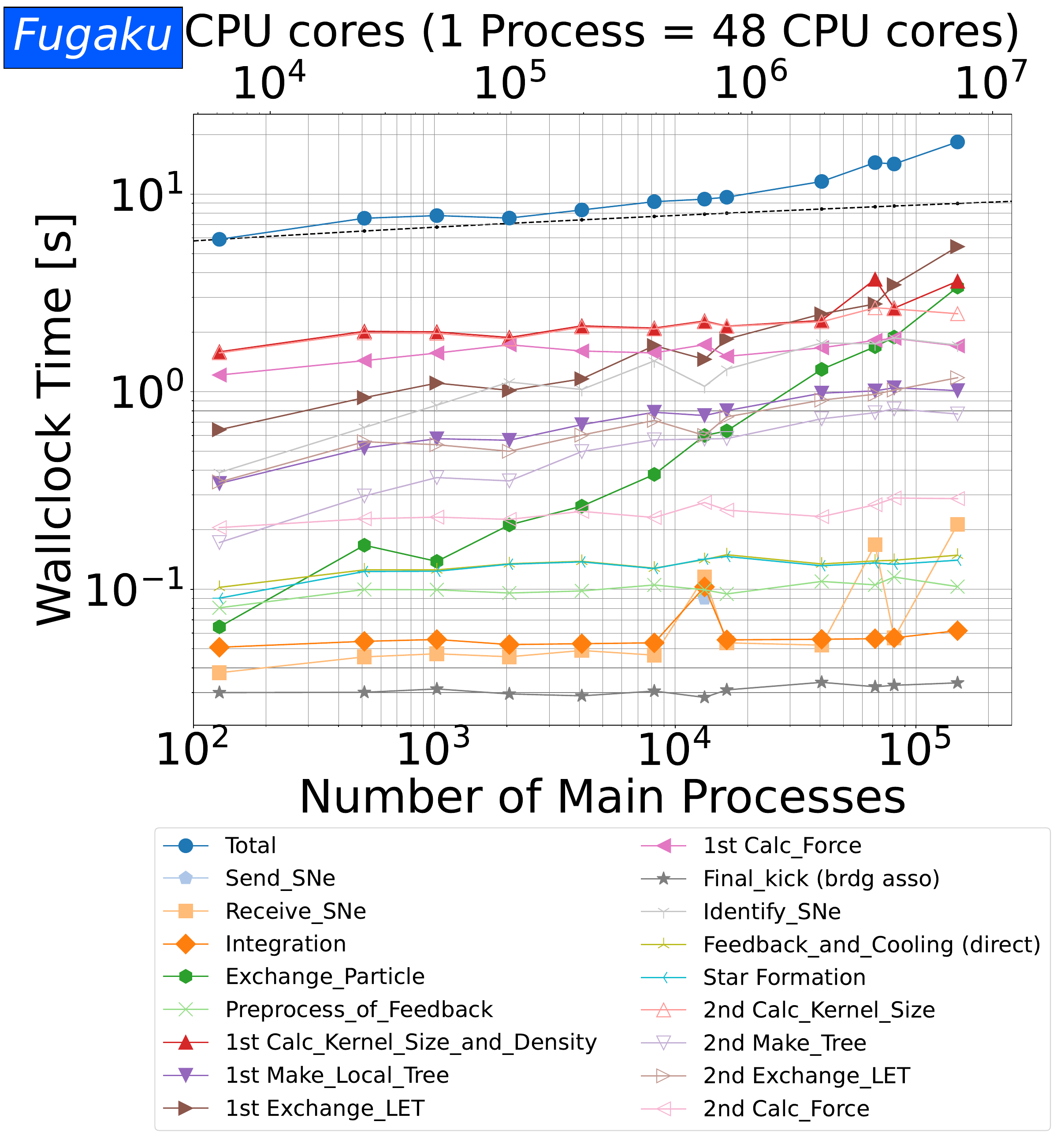}
\includegraphics[width=7.5cm]{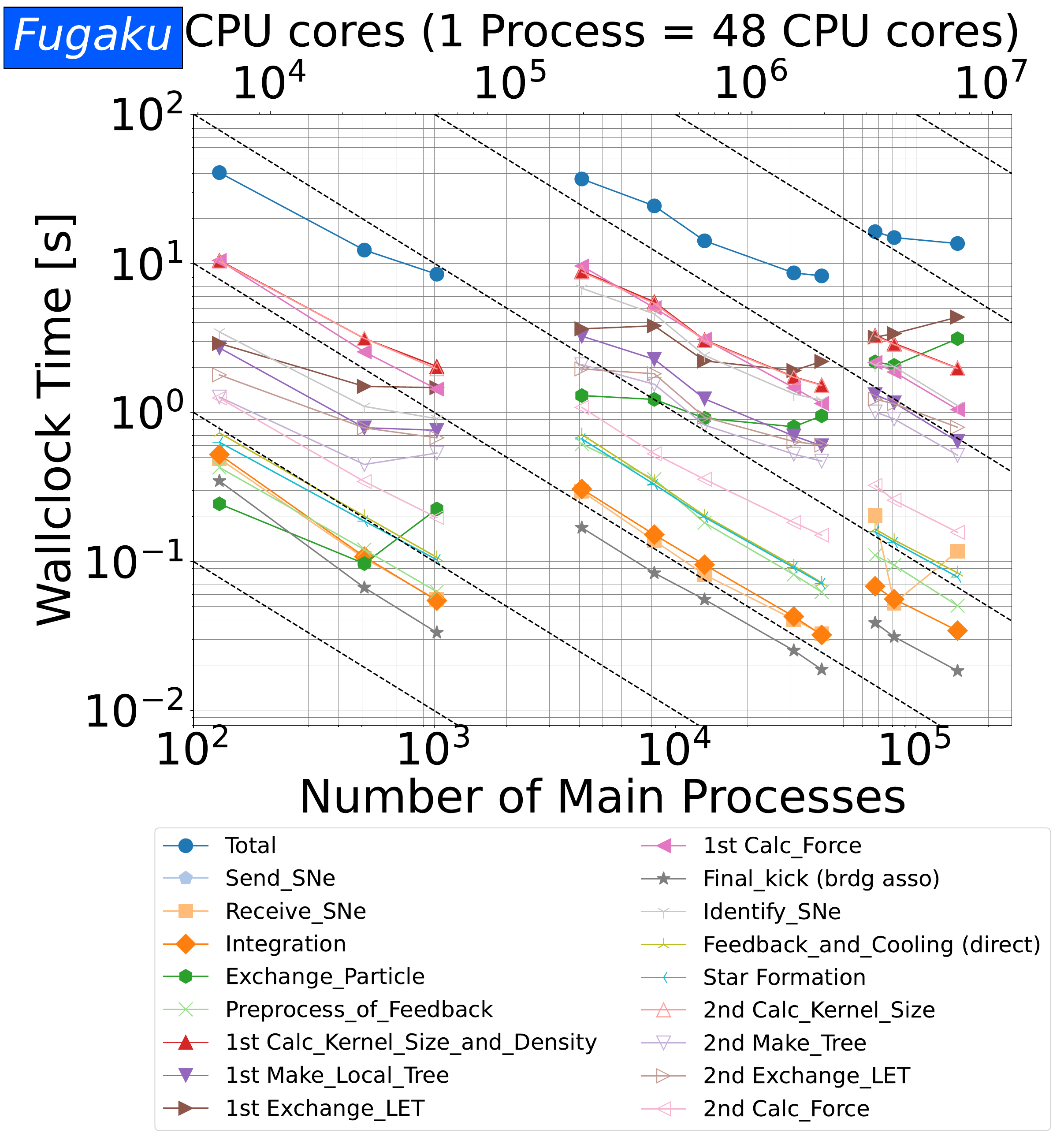}
\caption{(\textit{Left}) Weak-scaling performance: Wall-clock time per timestep. Each MPI process initially contains 2\,M particles, with one MPI process per compute node. Dashed line indicates $\propto \log N$. (\textit{Right}) Strong-scaling performance: Wall-clock time per timestep. Black dotted line shows ideal linear scaling. 
Total particle counts are $2.3\times 10^{10}$ and $1.5\times 10^{11}$, respectively.}
\label{fig:weak}
\vspace{-0.5cm}
\end{figure*}

We first show the weak-scaling performance of our code in Figure~\ref{fig:weak} measured on Fugaku. Here, the calculation time of `main nodes' is shown since the number of `pool nodes' is fixed, and the pool nodes work individually. We adopted our MW model and set the number of particles per node to be 2 million (2M). This value is limited by the memory size that we can use (32GB per node). 
We note that we fixed the galaxy size but changed the resolution to measure this weak scalability because it is challenging to scale up/down a single self-consistent galaxy model. As is also described in Section~\ref{sec:model}, the size of domains compared to the entire system (galaxy) becomes smaller as the number of MPI processes increases. 
We also note that the amount of calculations increases with $N\log N$, not $N$, where $N$ is the total number of particles. This is because the tree construction, traversal, and the size of the interaction list increase with $N\log N$. We, therefore, show a line $\propto \log N$ in Figure~\ref{fig:weak}. Considering the increase of the calculation cost with $\log N$, the efficiency of 148k nodes is 54\,\% of 128 nodes.

\begin{figure*}
\centering
\includegraphics[width=7.5cm]{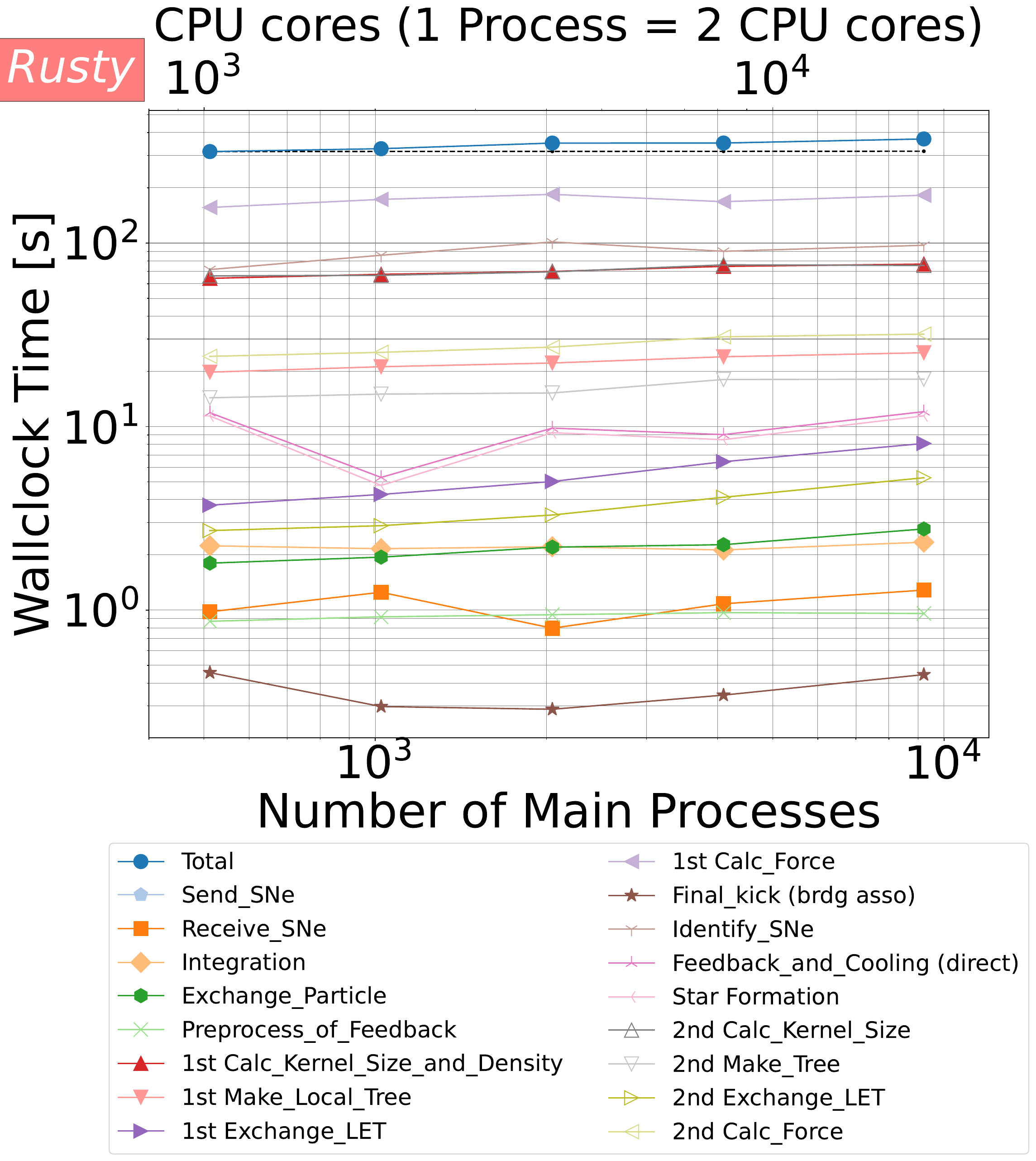}
\includegraphics[width=7.5cm]{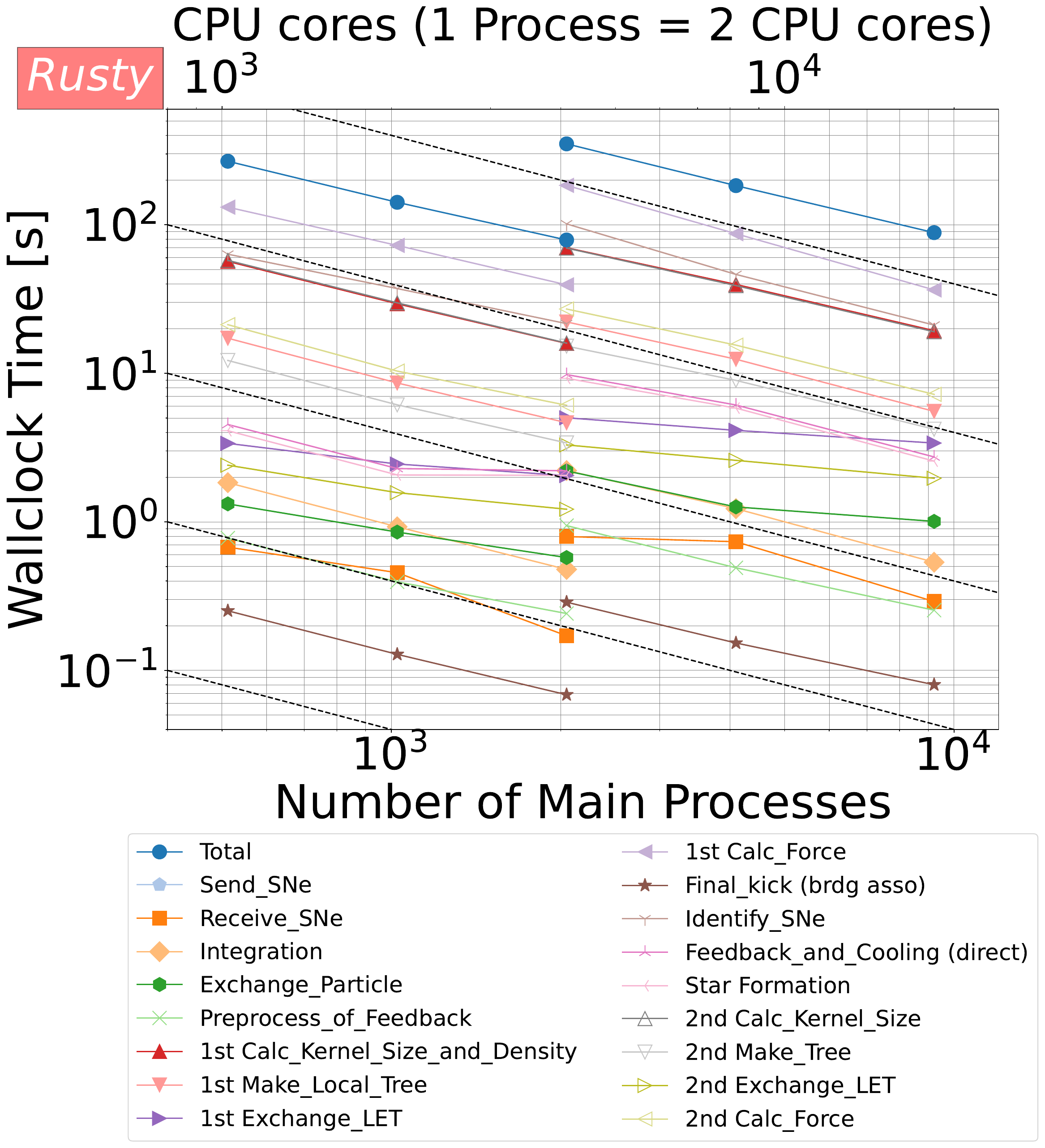}
\caption{(\textit{Left}) Weak-scaling performance: Wall-clock time per timestep on Rusty. Each MPI process starts with 25\,M particles, and 48 MPI processes are run per compute node. Dashed line again indicates $\propto \log N$. (\textit{Right}) Strong-scaling performance: Wall-clock time per timestep on Rusty. The black dotted line shows perfect linear scaling. 
Total particle counts of particles are $1.1\times 10^{10}$ and $5.1\times 10^{10}$, respectively.}
\label{fig:weak_rusty}
\vspace{-0.4cm}
\end{figure*}

Figure~\ref{fig:weak} presents the strong-scaling performance measured on Fugaku. Since the number of particles available on one node is limited, we adopted three different total numbers of particles for a small (128--1k), middle (4k--40k), and large (67k--148k) number of the main nodes (see Table~\ref{tab:runs}).  
The bottleneck calculation, such as interaction calculation (Calc Force) and Calc Kernel Size, scales very well. On the other hand, calculations requiring communications (Exchange LET and Exchange Particles) become a bottleneck as the number of MPI processes increases. 
The performance on Rusty (X86-64 processors) also shows excellent scalability, although the number of CPUs is an order-of-magnitude smaller than Fugaku (see Figures~\ref {fig:weak_rusty} for runs weakMW\_rusty and \ref{fig:weak_rusty} for runs strongMW\_rusty and strongMWs\_rusty listed in Table~\ref{tab:breakdown}).

The time for DL is not included here because it runs independently on the pool nodes and fully overlaps with this main integration part. The breakdown of the calculation time is summarized in section \ref{sec:breakdown}.

It is important to reach $\sim 10$\,sec per step. The timescale of galactic dynamics is $10^9$ year. If we adopt a fixed timestep of 2,000 years, the number of steps necessary for $10^9$ year integration is $5\times 10^5$. Assuming 10 sec per step, the calculation time is estimated to be $10\,[\mathrm{sec}] \times 5 \times 10^5 \sim 60$ days. 
This is reasonable for a single simulation.
We discuss more details in section \ref{sec:time-to-solution}.

\subsection{Anatomy of the performance} \label{sec:breakdown}
Table~\ref{tab:breakdown} shows the breakdown calculation time of run weakMW2M for 148900 (150k) nodes. 
The overall performance for one step was 8.20\,PF, and the efficiency was 0.90\%. 
The heaviest part of the calculation is the interaction calculation, especially for gravity. The performance of this part was 90.2\,PF, and the efficiency was 9.9\%. One may think the performance should be low, but we tuned the particle distribution to minimize the total calculation time, and therefore, the interaction calculation is tuned to minimize the sum of the gravity and hydro force. Therefore, the measurement of only the gravity shows an imbalance. In the following, we look more details to understand the performance.

\subsubsection{Exchange particles}
This part consists of two parts; determining a new domain for each node and exchanging particles following the new domains.  The domain decomposition requires communication among all MPI processes. We used an all-to-all scheme with $O(p^{1/3})$ as described in Sec.~\ref {sec:FDPS}. 

The particle exchange time increases as the number of MPI processes increases, and it was the second time-consuming part with the full system of Fugaku. This may be due to the shape of the domains. The data size increases as the surface of the domain increases. As shown in Figure~\ref{fig:DomainCoord}, some domain shows a long and thin structure. This shape increases the amount of particles to be exchanged and slows down this part.
We note that we do not have to decompose the domains and exchange particles every timestep, although we include them every timestep.

\subsubsection{Tree construction and walk}
The calculation cost of this part is of $O(N \log N_{\rm loc}/n_{\rm g})$, where $N_{\rm loc}$ is the number of particles per MPI process and $n_{\rm g}$ the average number of particles to share the interaction list. This part involves tree traversal, which requires random access to the main memory. Thus, this part requires high memory bandwidth for random access and also low latency. When we make $n_{\rm g}$ large, the calculation cost of this part decreases, but the calculation cost of the interaction
kernel increases. 

\subsubsection{LET exchange}
This part also requires an all-to-all communication because the gravitational force reaches the entire system. Because of the communication cost, this part is most time-consuming with the full system of Fugaku. 

\subsubsection{Interaction calculation}
This part requires the heaviest calculation. The calculation cost is $O(N \log N)$, where $N$ is the total number of particles. For more details, the performance of this part depends on the amount of memory access given by $O(N_{\rm loc}n_{\rm l}/n_{\rm g})$, where $n_{\rm l}$ is the average length of the interaction list, which is $O(\log N + n_{\rm g})$. On the other hand, the calculation cost is $O(N n_{\rm l})$. This means that as the necessary bytes per flop (B/F) also varies when we change $n_{\rm g}$, and the optimal choice for given hardware is necessary. 
We found $n_{\rm g}=2048$ best for Fugaku.  
As described above, the performance of the interaction calculation for gravity was 50.7\,PF, and the efficiency was 10\% for Fugaku using 81k nodes.

While we obtained the performance of Fugaku using a profiler, we had to measure the performance based on counting the calculations. For the other systems, therefore, we measured the performance of only the interaction kernels, for which we can easily count the number of calculations from the interaction counts. We obtained 0.863 and 0.209 PFLOPS for gravity and hydro force, respectively, using Rusty 193 nodes. From the scalability results, which scale well enough, we would be able to obtain a better performance using a similar but larger system, although we currently do not have access to a larger system. 
We also note that the number of particles in this test with the weakMW2M model on Rusty reached $2.3 \times 10^{11}$, which is approximately the same as the number of particles in the full system run on Fugaku.

For the GPU case, using nearly the entire Miyabi system (1024 nodes and 1024 GPUs), we measured the performance in gravity calculations, achieving 5.60 PFLOPS.
The efficiency was 8.1 \%. Currently, our code can utilize GPUs only for gravitational interactions, which are the bottleneck of this simulation (see a run MW\_miyabi listed in Table~\ref{tab:breakdown}). We found $n_\mathrm{g}$ = 65536 best for Miyabi.

\subsubsection{SPH kernel size}
This part includes both tree walk and interaction calculation, and they are repeated until the results converge. The iterations are usually twice, if we can set the initial guess of the kernel size properly. Every iteration requires communication with other MPI processes. In addition, the SPH kernel size strongly depends on the gas density. Some low density regions have a large SPH kernel size.

\begin{table}
\centering
\caption{Breakdown of calculation time and performance. }
\label{tab:breakdown}
\begin{tabular}{lrrr}
\hline
\multicolumn{4}{c}{Fugaku (A64FX) 150k nodes, (peak performance 915\,PFLOPS)} \\
\hline
Measured items                                       & Wall-time$^{\dagger}$ &  FLOP count  &  PFLOPS     \\
           & (sec) &  (PFLOP)  &  \\
\hline
Total time per step & 20.34 & $1.67\times 10^2$ & 8.20 \\
Particle exchange  & 3.87 & $3.57\times 10^{-8}$ & $9.25\times 10^{-9}$ \\
Tree construction  &  &  &   \\
\quad Gravity      & 0.96 & $1.25\times 10^{-2}$ & $1.31\times 10^{-2}$ \\
\quad Hydro Force  & 0.12 & $1.41\times 10^{-3}$ & $1.15\times 10^{-2}$ \\
LET Exchange       &  &  &    \\
\quad Gravity      & 3.89 & $1.26\times 10^{-2}$ & $3.25\times 10^{-3}$ \\
\quad Hydro Force  & 1.41 & $3.27\times 10^{-3}$ & $2.32\times 10^{-3}$ \\
Interaction calculations &  &  &  \\
\quad Gravity              & 1.63 & $1.47\times 10^2$ & 90.2 \\
\quad Hydro Force          & 0.34 & 4.36 & 13.0 \\
\quad Density and Pressure & 1.18 & 3.81 & 3.23 \\
Kernel Size Calculation    & 3.18 & 1.78 & 0.558 \\
\hline
\hline
\multicolumn{4}{c}{Rusty (genoa) 193 nodes, (peak performance 2.43\,PFLOPS)}\\
\hline
Measured items                                       & Wall-time &  FLOP count  &  PFLOPS     \\
           & (sec) &  (PFLOP)  &     \\
\hline
Interaction calculations &  &  &  \\
\quad Gravity       &  138 & 119 &  0.863\\
\quad Hydro Force  & 18.4 &  3.84 &  0.209 \\
\hline
\hline
\multicolumn{4}{c}{Miyabi (GH200) 1024 nodes, (peak performance 68.5 PFLOPS)}\\
\hline
Measured items    & Wall-time &  FLOP count  &  PFLOPS     \\
           & (sec) &  (PFLOP)  &     \\
\hline
Interaction calculations &  &  &  \\
    \quad Gravity      & 22.6 & 52.4 &  5.60\\
\hline
\end{tabular}
\begin{flushleft}
$^{\dagger}$ Shown are the elapsed time for the slowest MPI process for each item. \\
\end{flushleft}
\end{table}

\subsection{Time-to-Solution}\label{sec:time-to-solution}

In our models, we used a maximum of $3.0\times10^{11}$ particles. Since our timestep is fixed to 2,000 years, the timestep necessary for one million years is 500. The calculation time for one step is 20 seconds using 148,896 nodes, so the time-to-solution is 10,000 seconds (2.78 hours) for 1 million years.

We compare our time-to-solution to state-of-the-art conventional simulations, in which the timestep changes following the time evolution in the region (adaptive timestep). 
Because no performance data of the simulations listed in Table \ref{tab:pastsims_iso}, we use the data of GIZMO code \cite{Hopkins+2018} measured using a cosmological simulation. 
Their Figure G1 showed the performance of GIZMO code using a MW-size galaxy, which has a total galaxy mass similar to our model. 
From their figure, the CPU hours to integrate for $2\times10^9$ years were 0.05 million hours for using $1.5\times10^8$ particles. This figure also shows that the simulation does not speed up with more than 2,000 CPUs. Therefore, their fastest simulation took 0.0125 hours to integrate $1.5\times10^8$ particles for 1 million years. We need to consider an increase of timesteps, which follows $\propto N^{1/3}$, where $N$ is the number of particles; this increase is inevitable for conventional simulations using adaptive timesteps. 
Therefore, the necessary calculation time for 1 million years is estimated to be $(3\times10^{11}/1.5\times 10^{8})^{4/3} \times 0.0125=315$ hours for a 1 million year simulation. 
Thus,
our simulation is $113\times$ speedup compared to the current state-of-the-art simulation.

Another comparison to the current state-of-the-art simulations can be made using the number of timesteps. We performed simulations using our code without ML but with adaptive timesteps based on the CFL condition. We call it ``conventional simulation.''
The timestep of our conventional simulation shrank to 200 years after the SN, which is $10\times$ smaller than that adopted for the method with ML (2,000 yr). Thus, our code speeds up $10\times$ based on the timestep. The minimum timestep of the conventional simulations can be shortened even more after the galaxy structures have developed as the simulations proceed. Thus, our new method benefits more in the later stages of the simulation.

\subsection{Performance of interaction kernels}

In Table~\ref{tab:kernels}, 
we summarize the performance of interaction kernel calculations measured on single CPU cores and GPU card, which is the bottleneck of galaxy simulations, except for the timestep problem.
For the calculation of the gravitational interaction, our kernel of the monopole moment currently achieves an efficiency of 29.4\,\% on A64FX SVE as single-precision peak performance.
The A64FX processor has relatively high latency for floating-point arithmetic operations (e.g., 9 cycles for FMA), making loop unrolling necessary to achieve high computational efficiency.
However, the available number of architecture registers in the SVE instruction set of A64FX is not large enough to allow efficient loop unrolling\cite{Odajimaetal2018} to hide the large latency. We, therefore, divided the loops into small pieces (loop fission) to make the best use of the SIMD pipelines.
Because this loop fission brings the overhead of additional loads and stores of intermediate results and loop startup, the efficiency of A64FX is limited compared to that of the other architectures.

With AVX-512, the efficiency exceeded 60\% for all the kernels and was almost $70$\% for the gravity kernel.
The efficiency of AVX2 for the gravity kernel was 50.2\,\%, while that for the hydro kernels was only 22.4\,\% because of the table lookup. 
In AVX2 implementation, the gather load instruction is used for the table lookup, which may result in the relatively low performance of AVX2 hydro kernels. For ARM SVE and AVX-512, the table lookup works very well. 
We note that the theoretical peak performances with AVX2 and AVX-512 of the AMD EPYC™ 9474F are identical.

PIKG can be also used with GPUs. We measured the performance of a single GPU. The efficiencies of the gravity and hydro kernels using NVIDIA GH200 were 38.0\% and 2.8\%, respectively. This performance will be improved by tuning PIKG for GPUs.

\begin{table*}
\centering
  \caption{Asymptotic single core performance of interaction kernels using PIKG.}
  \label{tab:kernels}
  \resizebox{\textwidth}{!}{
  \begin{tabular}{lcccccccccc}
    \hline
Kernel   & \# of operations & Speed  & Efficiency & Speed  & Efficiency & Speed  & Efficiency & Speed  & Efficiency \\
    \hline
                      &     &  \multicolumn{2}{c}{Fugaku (A64FX SVE)} &  \multicolumn{2}{c}{Rusty (genoa AVX2)} &  \multicolumn{2}{c}{Rusty (genoa AVX512)} & \multicolumn{2}{c}{Miyabi (GH200)}\\
    \hline  
    Gravity & 27 & 37.7 Gflops & 29.4 \%  &  65.8 Gflops &  50.2 \% &  90.6 Gflops &  69.1 \% &  25.4 Tflops &  38.0 \% \\
    Hydro density/pressure & 73 &  21.9 Gflops & 17.1 \% & 15.1 Gflops &  11.5 \% &  87.6 Gflops &  66.8 \%  &  0.555 Tflops & 0.64 \% \\
    Hydro force & 101 & 19.8 Gflops &  15.4\% &  29.4 Gflops &  22.4  \% &  81.5 Gflops &  62.1 \% &  1.88 Tflops &  2.8 \% \\
\hline
\end{tabular}
}
\label{tab:PIKGgravity}
\end{table*}

\section{Implications}

We performed a galaxy simulation for the first time with the assistance of a DL surrogate model.
We demonstrated the performance of our highly efficient simulation code working with DL using 95\,\% of the full nodes (148,900 nodes) of Fugaku, which was available for our performance measurement.  

We also showed an excellent performance of our code using x86-64 CPU cluster.
The combination of FDPS and PIKG realizes both the portable high performance on diverse architectures, including x86-64 CPUs, ARM CPUs, and NVIDIA GPUs, and the high scalability from a single chip to the world-class supercomputers. In recent advancements of AI-specific accelerators as post-GPU computing, this approach helps us to utilize such new architectures with small porting effort.

Our novel integration scheme with a DL surrogate model enables us to adopt a constant shared timestep for all the particles. 
This allows us to perform massively parallel computation for galaxy simulations using $>$7,000,000 CPUs, which have been inefficient with previous methods. We achieved to utilize $\sim 500\times$ more particles and to $>100 \times$ speed up compared to the current-state-of-the art simulations.

The issue of small timestep is common in any high-resolution simulations, not only in galaxy simulations.
The technique of replacing a small part of simulations with DL surrogate models has the potential to bring benefits in various fields, especially in areas where it is essential to simultaneously simulate phenomena spanning both small and large scales or short and long time scales.
Similar methods to ours could be applied to simulations of cosmic large-scale structure formation, black hole accretion, as well as simulations of weather, climate, and turbulence.
The successful implementation of our novel DL-based approach marks a significant step forward in computational modeling, offering opportunities for enhanced efficiency and deeper insights into complex systems.

\begin{acks}

This research used computational resources of the supercomputer Fugaku at the RIKEN Center for Computational Science (Project ID: hp230204, hp240219, hp250226, hp250186), Miyabi-G awarded by "Large-scale HPC Challenge" Project, the Joint Center for Advanced High Performance Computing (JCAHPC), CfCA XC50 at National Astronomical Observatory of Japan, Wisteria/BDEC-01 at the University of Tokyo, and resources at the Flatiron Institute.
This study was partially supported by 
MEXT as “Program for Promoting Researches on the Supercomputer Fugaku”
(Grant Number JPMXP1020230406),
Research Organization for Information Science and Technology as the Advanced User-support Program (Full-node scale simulations on Fugkau), 
JSPS KAKENHI (21K03614, 21K03633, 22H01259, 22KJ0157, 22KJ1153, 22J11943, 23K03446, 24K07095, 25H00664, and 25K01046), JST FOREST Program (JPMJFR2367), Spanish grants PID2021-125451NA-I00 and CNS2022-135232 funded by \url{MICIU/AEI/10.13039/501100011033} and by ``ERDF A way of making Europe'', by the ``European Union'' and by the ``European Union Next Generation EU/PRTR'', and Initiative on Promotion of Supercomputing for Young or Women Researchers and Excellent Young Researcher Program of The University of Tokyo.
K.H. is financially supported by the JSPS (Research Fellowship for Young Scientists and Overseas Challenge Program for Young Researchers), JEES $\cdot$ Mitsubishi Corporation science technology student scholarship, and the IIW program of The University of Tokyo. K.H. also thanks CCA at the Flatiron Institute for hospitality while a portion of this research was carried out.
\end{acks}
\bibliographystyle{ACM-Reference-Format}
\bibliography{reference_min}
\end{document}